\documentclass[conference]{IEEEtran}
\IEEEoverridecommandlockouts
\usepackage{cite}
\usepackage{amsmath,amssymb,amsfonts}
\usepackage{algorithmic}
\usepackage{graphicx}
\usepackage{subcaption}
\usepackage{textcomp}
\usepackage{xcolor}
\usepackage{acronym}
\usepackage{hyperref}
\def\BibTeX{{\rm B\kern-.05em{\sc i\kern-.025em b}\kern-.08em
    T\kern-.1667em\lower.7ex\hbox{E}\kern-.125emX}}
    
\makeatletter
\def\endthebibliography{%
  \def\@noitemerr{\@latex@warning{Empty `thebibliography' environment}}%
  \endlist
}
\makeatother

\acrodef{ANN}{Artificial Neural Network}
\acrodefplural{ANNs}{Artificial Neural Networks}
\acrodef{CNN}{Convolutional Neural Network}
\acrodef{NN}{Neural Network}
\acrodefplural{NNs}{Neural Networks}
\acrodef{RNN}{Recurrent Neural Network}
\acrodefplural{RNNs}{Recurrent Neural Networks}
\acrodefplural{CNNs}{Convolutional Neural Networks}
\acrodef{DNN}{Deep Neural Network}
\acrodefplural{DNNs}{Deep Neural Networks}
\acrodef{SNN}{Spiking Neural Network}
\acrodefplural{SNNs}{Spiking Neural Networks}
\acrodef{GPU}{Graphical Processing Unit}
\acrodefplural{GPUs}{Graphical Processing Units}
\acrodef{IoT}{Internet of Things}
\acrodef{FC}{Fully-Connected} 
\acrodef{FF}{Feed-Forward}
\acrodef{FFNN}{Feed-Forward Neural Network}
\acrodefplural{FFNNs}{Feed-Forward Neural Networks}
\acrodef{ReLU}{Rectified Linear Unit}
\acrodef{ML}{Machine Learning}
\acrodef{LIF}{Leaky Integrate and Fire}
\acrodef{IF}{Integrate and Fire}
\acrodef{RC}{Reservoir Computing}
\acrodef{STDP}{Spike Timing Dependent Plasticity}
\acrodef{PNN}{Photonic Neural Network}
\acrodefplural{PNNs}{Photonic Neural Networks}
\acrodef{BPTT}{Back-Propagation Through Time}
\acrodef{SOI}{Silicon-on-Insulator}
\acrodef{SiNOI}{Silicon Nitride-on-Insulator}
\acrodef{FDSOI}{Fully-Depleted Silicon-On-Insulator}
\acrodef{MZI}{Mach-Zenhder Interferometer}
\acrodefplural{MZIs}{Mach-Zenhder Interferometers}
\acrodef{RR}{Ring Resonator}
\acrodef{PCM}{Phase Change Material}
\acrodefplural{PCMs}{Phase Change Materials}
\acrodef{FLOP}{Floating Point Operations per Second}
\acrodef{MLP}{Multilayer Perceptron}
\acrodef{RTL}{Register-Transfer Level}
\acrodef{ST}{Shadow Training}
\acrodef{ASIC}{Application Specific Integrated Circuit}
\acrodefplural{ASICs}{Application Specific Integrated Circuits}
\acrodef{FPGA}{Field Programmable Gate Arrays}
\acrodef{AER}{Address Event Representation}
\acrodef{GSOPS}{Giga-Synapse Operations}
\acrodef{SGD}{Stochastic Gradient Descent}

\begin{document}
\bstctlcite{IEEEexample:BSTcontrol}
\IEEEoverridecommandlockouts
\IEEEpubid{\makebox[\columnwidth]{ 979-8-3503-4630-5/23/\$31.00 \copyright2023 IEEE \hfill} \hspace{\columnsep}\makebox[\columnwidth]{ }}

\title{Special Session: Neuromorphic hardware design and reliability from traditional CMOS to emerging technologies
\thanks{This project has received funding from the European Union’s Horizon Europe research and innovation programme under grant agreement No. 101070238. Views and opinions expressed are however those of the author(s) only and do not necessarily reflect those of the European Union. Neither the European Union nor the granting authority can be held responsible for them.
\\It was also partially supported by the ANR within the EMINENT Project ANR-19-CE24-0001}
}

\author{\IEEEauthorblockN{Fabio~Pavanello$^1$, Elena~Ioana~Vatajelu$^2$, Alberto~Bosio$^1$, Thomas~Van~Vaerenbergh$^3$, Peter~Bienstman$^4$,\\Benoit~Charbonnier$^5$, Alessio~Carpegna$^6$, Stefano~Di~Carlo$^6$, Alessandro~Savino$^6$}
\IEEEauthorblockA{$^1$Univ. Lyon, Ecole Centrale de Lyon, INSA Lyon, Université Claude Bernard Lyon 1, CPE Lyon, CNRS, INL\\
$^2$Univ. Grenoble Alpes, CNRS, Grenoble INP, TIMA, 38000 Grenoble, France\\
$^3$Hewlett Packard Labs, HPE Belgium, B-1831 Diegem, Belgium, 
$^4$Ghent University - imec, Gent, Belgium\\
%$^5$Univ. Grenoble Alpes, CEA, LETI, Grenoble, France\\
$^5$CEA-LETI, Grenoble, France\\
$^6$Politecnico di Torino, Control and Computer Eng. Department, Torino, Italy\\
%Emails: fabio.pavanello@cnrs.fr, ioana.vatajelu@univ-grenoble-alpes.fr, alberto.bosio@ec-lyon.fr, \\ thomas.van-vaerenbergh@hpe.com, peter.bienstman@ugent.be, \{alessio.carpegna, stefano.dicarlo, alessandro.savino\}@polito.it
}
}

\maketitle

\begin{abstract}
The field of neuromorphic computing has been rapidly evolving in recent years, with an increasing focus on hardware design and reliability. This special session paper provides an overview of the recent developments in neuromorphic computing, focusing on hardware design and reliability. We first review the traditional CMOS-based approaches to neuromorphic hardware design and identify the challenges related to scalability, latency, and power consumption. We then investigate alternative approaches based on emerging technologies, specifically integrated photonics approaches within the NEUROPULS project. Finally, we examine the impact of device variability and aging on the reliability of neuromorphic hardware and present techniques for mitigating these effects. This review is intended to serve as a valuable resource for researchers and practitioners in neuromorphic computing.
\end{abstract}


\section{Introduction}
%Intro to neuromorphic hardware and spiking neural network

\acp{ANN} have enabled complex computations but require significant computational resources for both training and inference \cite{Marchisio:2019aa}. The main bottleneck in these networks is the transfer of large amounts of data to support different tasks. However, the trend in computing is moving towards edge devices, such as the \ac{IoT}, to improve security and reduce power consumption and latency \cite{Shafique:2018aa}. Furthermore, there is a growing demand for powerful yet energy-efficient accelerators in various fields, including fault detection in microprocessors \cite{Dutto:2021aa} and intrusion detection systems \cite{Sedjelmaci:2016aa}. 

The design space for ANNs is vast, and it involves choices in four fundamental aspects: the neuron model, the architecture structure, the information encoding, and the training method \cite{zhou2018brief}. These choices can significantly impact the hardware design and optimization process. \acp{CNN}, \acp{DNN}, and \acp{SNN} are three popular types of \acp{ANN}, each with unique characteristics and applications.

\acp{CNN} are commonly used in image recognition and processing tasks, and they rely on convolutional layers to extract spatial features from input images \cite{LeCun1998}. \acp{DNN}, on the other hand, are used in a wide range of applications, from speech recognition to natural language processing and game playing \cite{goodfellow2016deep}. They are characterized by multiple layers of neurons that learn increasingly abstract features of the input data.

The limitations of traditional computing architectures, particularly the communication bottleneck between memory and processor and the latency of information propagation and manipulation, have highlighted the need for alternative approaches to \acp{ANN}. While software approaches for \acp{ANN} offer advantages when implemented on specialized hardware such as \acp{GPU}, these limitations persist~\cite{schuman_opportunities_2022}. 

\acp{SNN} have emerged as the next generation of \acp{ANN} that exchange information in spikes, inspired by the behavior of biological brains \cite{ThirdGen}. This allows for more efficient computation and reduced power consumption and is particularly interesting when working with time sequences such as audio, video, and electrical signals \cite{Heildelberg, EcgSnn}. The model complexity and internal parameters determine the model's suitability to the input data, with shorter time constants detecting shorter temporal correlations and higher values catching more prolonged time effects.

Eventually, the resilience of the \ac{ANN} is crucial when designing the entire system and cannot be ignored~\cite{Vallero:2019aa,Ruospo:2023aa}. Retaking inspiration from biology, the human brain is intrinsically resilient to malfunctioning and faults. It loses approximately 50000 neurons daily but can still perform complex tasks and learn new ones, creating connections between the remaining neurons. \acp{ANN} have inherited this characteristic at a certain level, but they still need to improve significantly, particularly in mission-critical and safety-critical applications. Therefore, a deeper study of \acp{ANN} from this perspective is required.

Neuromorphic computing, which merges memory and processing units within neurons and synapses and maps computing architectures more closely to \acp{NN} models, offers a promising solution to address the limitations of traditional computing architectures \cite{james_historical_2017}. Various technologies, including CMOS, memristors, and optoelectronics/all-optical approaches, have been explored to develop neuromorphic hardware. This paper provides an overview of these approaches, highlighting their advantages and limitations. Background on \acp{ANN} is given in \autoref{sec:background}, while \autoref{sec:digital} discusses the design and performance of the digital version of \acp{ANN}, with a focus on the \acp{SNN}. \autoref{sec:photonics} discusses the potential of silicon photonics for \acp{ANN}, and \autoref{sec:reliability} covers reliability challenges in the usage of \acp{ANN}.

\section{Background}
\label{sec:background}

The choice of the neuron model, architecture structure, information encoding, and training method can significantly impact the hardware design and optimization process for each of these \acp{ANN}. Therefore, it is essential to consider these aspects carefully when designing and implementing hardware for \acp{ANN}.

\acp{CNN} and \acp{DNN} use different neuron models. In \acp{CNN}, the neuron model is typically based on the \ac{ReLU} function, a non-linear activation function commonly used in deep learning. The \ac{ReLU} function is simple and computationally efficient, making it a popular choice for \acp{CNN}. The \ac{ReLU} function is $f(x) = \max(0,x)$, where $x$ is the input to the neuron. The output of the \ac{ReLU} function is zero if the input is negative and equal to the input if the input is positive. The \ac{ReLU} function effectively reduces overfitting in \acp{DNN} and has been used in various computer vision tasks, such as object detection and recognition \cite{Krizhevsky2012}.

In \acp{DNN}, the neuron model is typically based on a non-linear activation function. The sigmoid function, defined as $f(x) = \frac{1}{1+e^{-x}}$, where $x$ is the input to the neuron, is one of the simplest. The output of the sigmoid function is between 0 and 1, which makes it useful for tasks such as classification. Another popular activation function in \acp{DNN} is the hyperbolic tangent ($\tanh$) function, similar to the sigmoid function but outputs values between -1 and 1. The choice of activation function depends on the task at hand and the \acp{NN}'s architecture. For example, the sigmoid and $\tanh$ functions were commonly used in early \ac{DNN} architectures, such as the \ac{MLP} \cite{rumelhart1986learning} but have since been largely replaced by the \ac{ReLU} function in more recent architectures. However, the sigmoid and $\tanh$ functions are still used in certain NNs, such as \acp{RNN} and autoencoders \cite{Hinton:2006aa}.

State-of-the-art implementations of \acp{CNN} and \acp{DNN} often use 32-bit floating-point numbers in software or model-based approaches. However, implementing such algorithms in hardware is challenging due to their extensive data requirements, high energy consumption, and large memory bandwidth. To address these challenges, quantization, and regularization techniques have been explored, using fixed-point computations with 16 bits, 8 bits, or lower precision. Although these methods reduce the precision of synaptic weights and inter-layer signals, IBM's TrueNorth~\cite{TrueNorth} chip has achieved acceptable precision using only five synaptic states, albeit at high design and energy costs \cite{DRAGHICI2002395, doi:10.1126/science.1254642}.

In the case of \acp{SNN}, the neuron models and architectures are the most complex to target due to the nature of the information they carry and how they treat it. Many different mathematical models describe and mimic the behavior of biological neurons, such as the Hodgkin-Huxley model \cite{HodgkinHuxley}, the Izhikevich model \cite{Izhikevich}, the \ac{LIF} model \cite{Glif}, and the \ac{IF} model \cite{IfReview}. These models range from very complex and detailed to much simpler and more suitable for machine learning and hardware applications, with varying degrees of biological plausibility and computational efficiency.

Neurons in \acp{SNN} are generally treated as leaky integrators where input spikes are integrated over time after being weighted by corresponding synapses, affecting the neuron's state, usually the electrical potential across its membrane. The neuron membrane depolarizes due to internal charge leakage without spikes, except in the \ac{IF} model, where it is kept at a constant value. A new spike is generated when the membrane potential exceeds a specific threshold value, causing the neuron to fire, and the potential drops suddenly into a reset state.

The architecture of the \ac{SNN} can also be very flexible. The literature reports \ac{FC} \ac{FF} \ac{SNN}~\cite{Carpegna:2022aa}, regularly recurrent structures~\cite{Heildelberg} or randomly recurrent architectures~\cite{SnnReservoir}, used for example, in \ac{RC}. There can be only excitatory connections, with positive weights, or adding inhibitory connections, with negative weights~\cite{Carpegna:2022aa}, or in more detailed models, separated excitatory and inhibitory neurons, as observed in some regions of the human brain.

\ac{RC} is an efficient technique where a randomly initialized \ac{RNN} trains only a linear combination of the signals at each node. It has shown promising results in various applications, including photonics \cite{vandersandeAdvancesPhotonicReservoir2017}. Different \ac{RC} architectures have been investigated using photonics, such as spatial-multiplexing and time-multiplexing approaches. Spatial-multiplexing involves physically separating the nodes, while time multiplexing requires a faster sampling speed and more complex processing of the read-out layer.

The choice of input encoding can significantly affect the performance of the \ac{CNN} or \ac{DNN}. It often requires careful consideration and experimentation to determine the optimal encoding for the given task. Factors to consider when selecting an input encoding include the nature and complexity of the input data, the available computational resources, and the application's performance requirements.

The most common input encoding for \acp{CNN} is raw pixel values, where each pixel in the image is represented as a numerical value. In contrast, for \acp{DNN} that process non-image data, input encoding may involve feature engineering techniques such as transforming the raw input data into a set of meaningful features more amenable to learning by the network~\cite{Krizhevsky2012}. For example, in natural language processing, input encoding may involve converting text into a numerical representation, such as bag-of-words or word embeddings.
 
Due to their nature, \acp{SNN} require more advanced methods for encoding and interpreting information. The main approaches are rate coding~\cite{RateCoding}, temporal coding\cite{TemporalCoding}, and population rank coding\cite{PopulationCoding}. Rate coding uses the average spike frequency to encode information and is suitable for static input data. It is less efficient regarding spike activity but more robust to noise. Temporal coding encodes information in the precise arrival time of spikes or their relative distance, requiring fewer spikes to process information, and is suitable for encoding time-varying signals. However, it is more sensitive to noise. Population rank coding uses the joint activity of a group of neurons to process information.

Training approaches involve optimizing the model parameters to minimize a given loss function for \acp{CNN} and \acp{DNN}. In supervised learning, this involves iteratively adjusting the weights and biases of the network to reduce the difference between the predicted output and the ground truth labels. The optimization is typically performed using gradient descent methods, which involve calculating the gradient of the loss function concerning the model parameters and updating them in the direction of the negative gradient. Commonly used gradient descent methods include \ac{SGD}~\cite{SGD}, AdaGrad~\cite{AdaGrad}, etc.

Regularization techniques such as Dropout~\cite{Dropout}, etc., are often used to prevent overfitting and improve generalization. Additionally, data augmentation methods such as flipping, rotating, and cropping the input images are employed to increase the size of the training dataset and improve model robustness~\cite{Takahashi:2018aa}.

Unsupervised learning approaches, such as Autoencoders~\cite{Hinton:2006aa} and Restricted Boltzmann Machines~\cite{Fischer:2012aa}, can also be used for pretraining the model parameters. Transfer learning approaches can also be employed, where a pre-trained network is fine-tuned on a new dataset with similar or related features~\cite{TransferLearning}. Finally, reinforcement learning can also be used to train \acp{CNN} and \acp{DNN}, where the model learns to take actions based on a reward signal, such as in game-playing agents~\cite{RL}.

Training \acp{SNN} is challenging due to the non-differentiability of the thresholding function of neurons. Classical back-propagation methods cannot be directly applied, but several approaches have been developed, including supervised and unsupervised training. In supervised training, the most common approach is to convert an \ac{ANN} into an \ac{SNN}, where the \ac{ANN}'s differentiable non-linear function is trained using back-propagation, and the weights are used directly in the \ac{SNN}. Alternatively, a \ac{BPTT} can be applied directly to the \ac{SNN}, where a surrogate gradient replaces the neuron's thresholding function with a differentiable function during the backward pass. In contrast, inspired by biology, most unsupervised approaches update weights locally based on the relative spike timing between the inputs and the output without depending on a global error signal propagating across the network, resulting in a lighter memory footprint and computational overhead, with \ac{STDP} being the most common method~\cite{BiPooStdp}.

\section{Digital accelerators}
\label{sec:digital}

As seen before, the architecture of an \ac{ANN}, and in the same way of an \ac{SNN}, is generally composed of many independent neurons and, as such, is intrinsically strongly parallelizable. This poorly fits the common CPU-based computing approach, in which the parallelism is limited to a few tens of very powerful cores. For this reason, one active research branch in the field of \ac{ANN} and \ac{SNN} is directed towards accelerating such algorithms, using computing platforms to execute them more efficiently. The goal is to broaden their application to many contexts, such as performance-constrained, power-constrained, or real-time tasks.

\begin{figure}[ht!]
         \centering
         \begin{subfigure}{0.6\columnwidth}
             \centering
             \includegraphics[width=\textwidth]{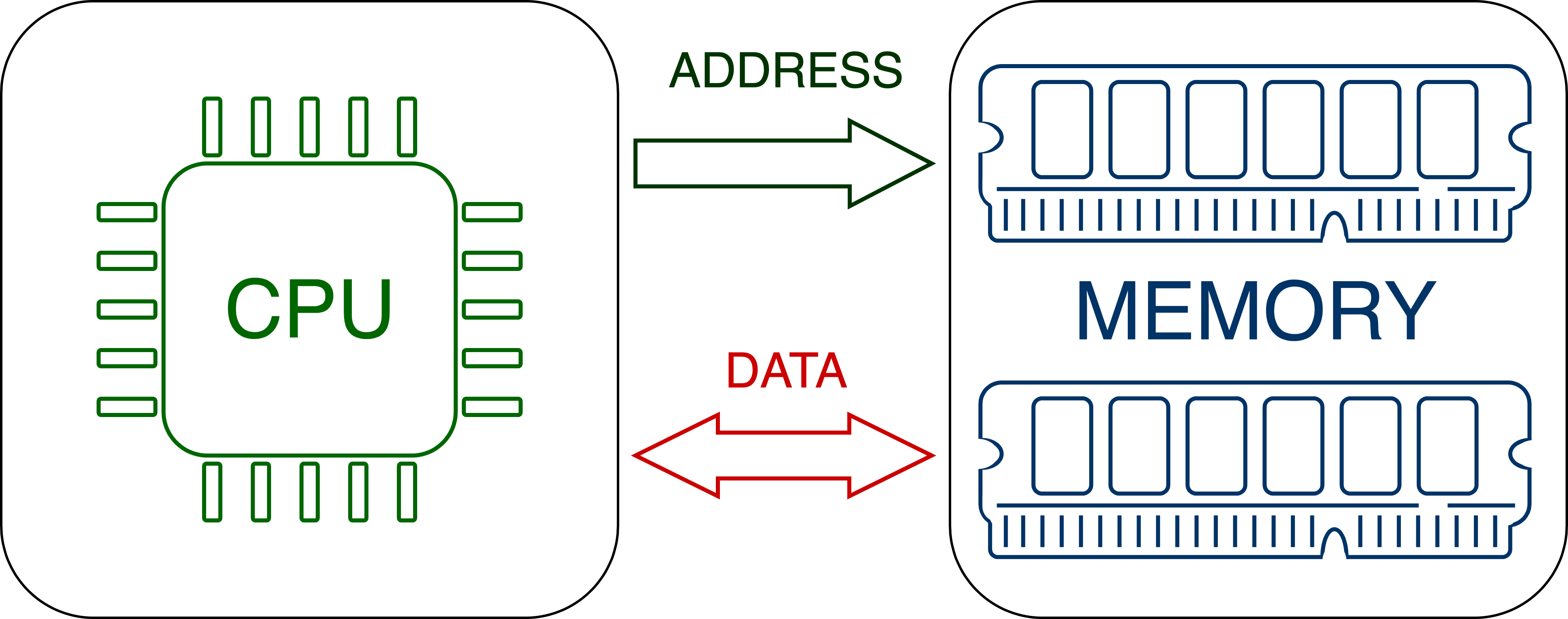}
             \caption{Classical Von Neuman computing approach}
             \label{fig:von_neuman}
         \end{subfigure}

         \hfill
         
         \begin{subfigure}{0.5\columnwidth}
             \centering
             \includegraphics[width=\textwidth]{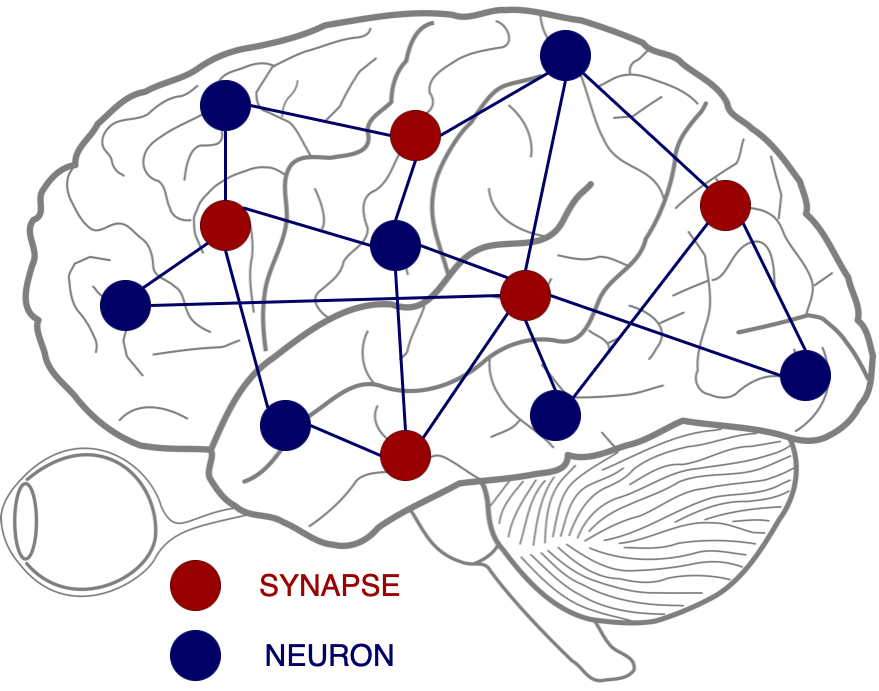}
             \caption{Neural network's computation and memory co-location}
             \label{fig:ann}
         \end{subfigure}

            %\caption{Different computing paradigm}
            \label{fig:vonNeuman_vs_ANN}
    \end{figure}

When discussing acceleration in the digital domain, several solutions have different degrees of optimization, efficiency, and cost. One first approach, already mainstream for many \ac{ANN} models, is to exploit the computational parallelism offered by general-purpose hardware accelerators like \acp{GPU}. Regarding \acp{SNN}, several software frameworks natively support the deployment of the code on \acp{GPU}, for example, the ones based on \emph{pyTorch}, like \emph{snnTorch}\cite{SnnTorch} and \emph{spikeTorch}, or CUDA accelerated C++ frameworks, like \emph{SLAYER}\cite{Slayer} and CARLsim 4\cite{Carlsim}.

However, \acp{SNN} have many features unsuitable with a \ac{GPU} execution. For example, spikes can be represented in the digital domain as single-bit events (high in the presence of a spike and low otherwise). The numerical representation used in \ac{GPU}, based on words with 8, 16, 32, 64, or similar bit widths, is inefficient. Moreover, the computation can be based on events: a neuron can react only in the presence of an active input spike, remaining in a quiescent state otherwise. Such an update policy would allow exploiting the sparsity of spikes typical of \acp{SNN}, with considerable savings in switching power, but again is not supported by \acp{GPU}.

This incompatibility has pushed for developing specialized hardware accelerators explicitly designed to support \ac{SNN} features. Interestingly \acp{SNN} are intrinsically more suitable for developing these accelerators than other \ac{ANN} models. The main reason is again how \ac{SNN} encodes the information, which in the digital domain corresponds to single-bit signals. This drastically reduces the interconnection and memory requirements.

    \begin{figure}[ht]
         \centering
         \begin{subfigure}{0.7\columnwidth}
             \centering
             \includegraphics[width=\textwidth]{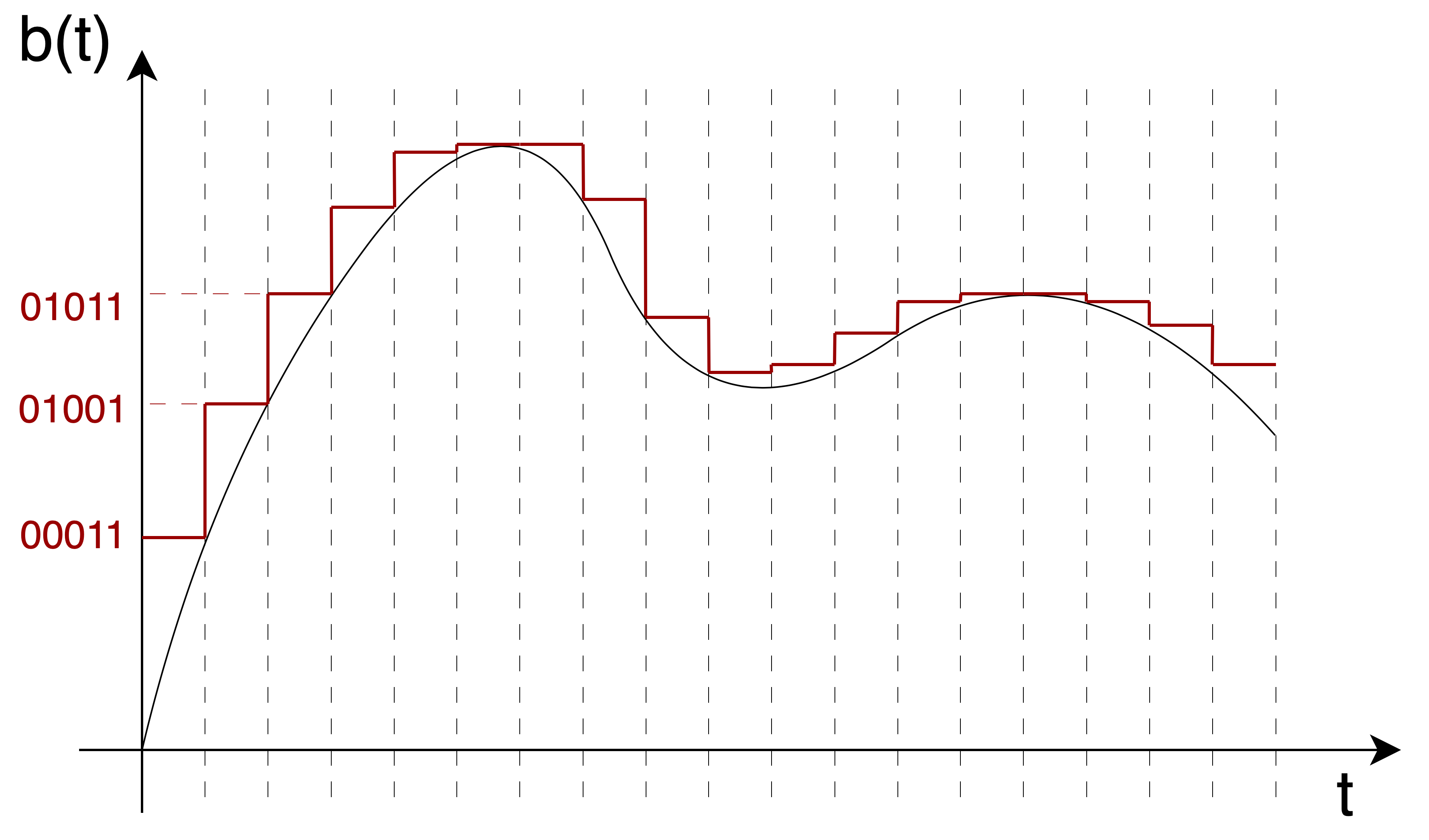}
             \caption{Multi-bit digital coding}
             \label{fig:digital}
         \end{subfigure}
         
         \hfill
         
         \begin{subfigure}{0.8\columnwidth}
             \centering
             \includegraphics[width=\textwidth]{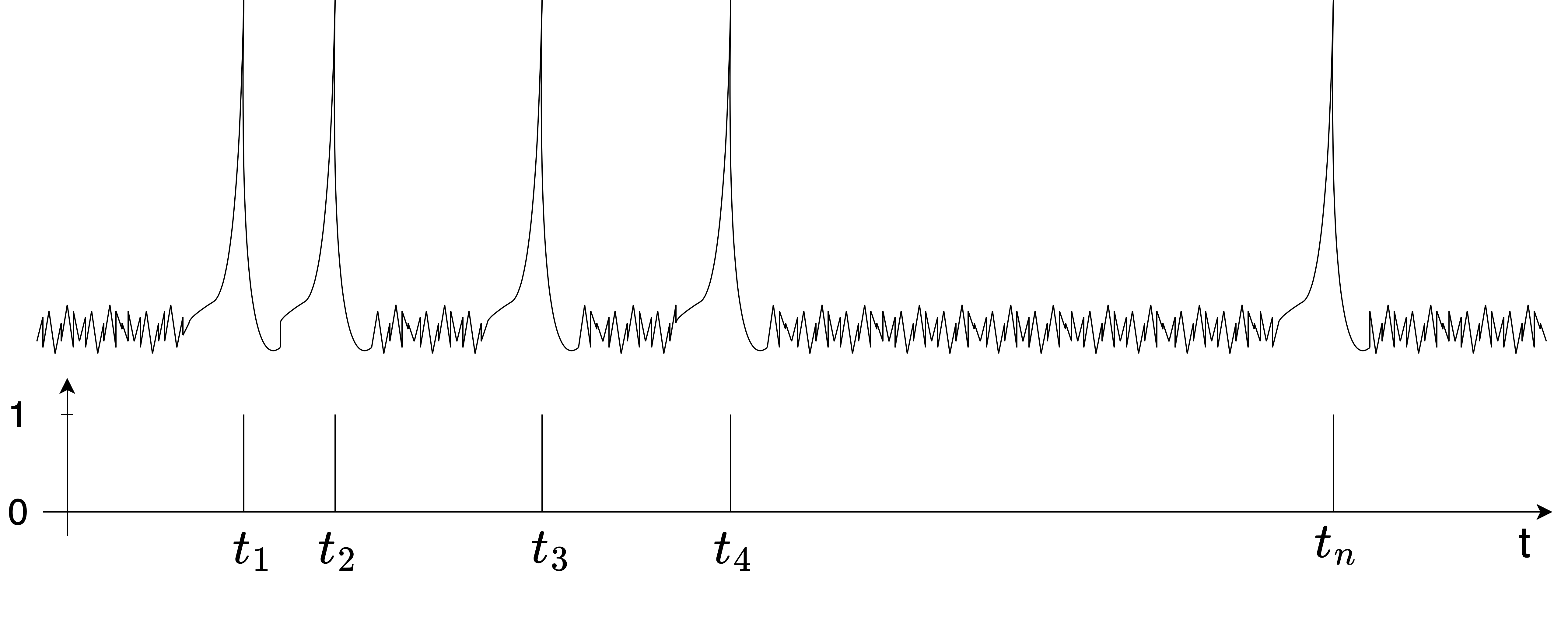}
             \caption{Spikes information encoding}
             \label{fig:spikes}
         \end{subfigure}
            %\caption{Different information representation}
            \label{fig:digital_vs_spiking}
    \end{figure}

When designing hardware accelerators, the roads that can be followed are two: fixed hardware or reconfigurable hardware. In the first case, the accelerator becomes an \ac{ASIC}, able to perform in a very efficient way the specific tasks for which it has been designed, but nothing more. On the other hand, specific hardware platforms can be reconfigured multiple times, allowing for flexible architectures. This is the case of \ac{FPGA}. The choice between the two depends a lot on the kind of application and constraints. Generally, designers try to fit as many functionalities as possible when developing an \ac{ASIC} since the hardware remains fixed. 

Companies and universities are investing in developing programmable chips to enable fast simulation of large-scale \ac{SNN}. Among them, a first attempt is SpiNNaker~\cite{SpiNNaker}. The idea behind SpiNNaker is to use classical CMOS architectures, particularly ARM968 RISC processors, to simulate the neurons' dynamics and optimize the routing between them to fit a spike-based communication perfectly. A specific communication protocol, i.e., \ac{AER}, is used to do this. This protocol is designed explicitly for neuromorphic circuits, in which a spike is represented through the ID of the neuron that generated it and the corresponding generation time stamp. The IBM TrueNorth~\cite{TrueNorth} is a fully custom \ac{ASIC} realized in 28nm CMOS technology. Again the neurons use standard CMOS digital gates, but the communication is asynchronous. The Intel Loihi~\cite{Loihi} has an approach that is a hybrid between the previous two. It has 128 custom chips implementing 1024 \ac{LIF} primitives. The custom connection mesh optimizes the typical sparse communication between spiking neurons. Additionally, Loihi includes specific components to perform learning directly on-chip. It provides a set of configurable parameters to allow different local learning rules, from a basic \ac{STDP} to more complex alternatives. Finally, Tianjic~\cite{Tianjic} is a hybrid \ac{ANN}/\ac{SNN} accelerator with a custom interconnection between the various cores.

In the field of \acp{ASIC}, it is worth citing ODIN~\cite{Odin}, a small network implemented in 28nm \ac{FDSOI} CMOS technology and targeting low-power applications. Neurons can be configured to implement a LIF model or one of the 20 possible Izhikevich behaviors. The routing of the spikes between neurons is again performed through an AER protocol. 

Table \ref{tab:ASICcomparisonTable} compares the accelerators in terms of area and energy efficiency, expressed in \ac{GSOPS} per Watt. For a more detailed comparison, see \cite{SNN_HW_review}.

\begin{table}[htb]

        \caption{ASIC comparison table\cite{SNN_HW_review}}

        \label{tab:ASICcomparisonTable}

        \centering

        \begin{tabular}{|c|c|c|c|c|c|c|}

            \hline

            \textbf{Design}             & 
            \cite{SpiNNaker}   &
            \cite{TrueNorth}   &
            \cite{Loihi}			&
            \cite{Tianjic}		&
            \cite{Odin}             \\

            \hline

            \textbf{\# of neurons}	& 
            18k				&
            1M  			&
            128k			&
            39k				&
            256 			\\

            \hline

            \textbf{\# of synapses}		&
            18M			&
            256M		&
            128M		&
            9.75M		&
            64k		\\

            \hline

            \textbf{Area ($mm^2$)}	&
            88.4		&
            413			&
            60			&
            14.4        &
            0.086       \\

            \hline

            \textbf{Process ($nm$)}		&
            130				&
            28				&
            14				&
            28				&
            65				\\

            \hline

            \textbf{Energy (GSOPS/W)}		&
            0.033			&
            400 			&
            -   			&
            649	    		&
            78.7			\\

            \hline

        \end{tabular}
        
    \end{table}

 \begin{table*}[htb]

        \caption{FPGA comparison table\cite{sixuLi}}

        \label{tab:FPGAcomparisonTable}

        \centering

        \begin{tabular}{|c|c|c|c|c|c|}

            \hline

            \textbf{Design}			& 
            \cite{minitaur}			&
            \cite{wangQian2017Eepn}		&
            \cite{darwin}			&
            \cite{sixuLi}			&
            \cite{Carpegna:2022aa}		\\

            \hline

            \textbf{Clock frequency(MHz)}	& 
            75				&
            120				&
            25				&
            100				&
            100				\\

            \hline

            \textbf{Data format}		&
            16bit Fixed			&
            8bit Fixed			&
            32bit Fixed			&
            16bit Floating		&
            16bit Fixed			\\

            \hline

            \textbf{Computing scheme}	&
            Event-Driven			&
            Clock-Driven			&
            Event-Driven			&
            Adaptive Clock/Event-Driven	&
            Clock-Driven			\\

            \hline

            \textbf{Neuron model}		&
            LIF				&
            LIF				&
            LIF				&
            LIF				&
            LIF				\\

            \hline

            \textbf{FPGA platform}		&
            Spartan 6			&
            Virtex 6			&
            Spartan 6			&
            Virtex 7			&
            Artix 7				\\

            \hline

            \textbf{Neurons}		&
            1794				&
            1591				&
            1794				&
            1094				&
            1384				\\

            \hline

            \textbf{Synapses}		&
            647000				&
            638208				&
            647000				&
            177800				&
            313600				\\

            \hline

            \textbf{Task}			&
            MNIST				&
            MNIST				&
            MNIST				&
            MNIST				&
            MNIST				\\

            \hline

            \textbf{Computation time}	&
            $0.53s$/image			&
            $8.40s$/image			&
            $0.16s$/image			&
            $3.15ms$/image			&
            $215\mu s$/image		\\

            \hline
            
            \textbf{Computation time @100MHz}	&
            $0.40s$/image			&
            $10.08s$/image			&
            $40.00ms$/image			&
            $3.15ms$/image			&
            $215\mu s$/image		\\

            \hline

            \textbf{Energy}			&
            $0.80J$/image			&
            $1.12J$/image			&
            Not reported			&
            $5.04mJ$/image			&
            $13mJ$/image			\\

            \hline

            \textbf{Energy/Synapse}		&
            $1.2\mu J$/synapse		&
            $1.76\mu J$/synapse		&
            Not reported			&
            $0.028\mu J$/synapse		&
            $0.041\mu J$/synapse		\\
            
            \hline

        \end{tabular}
        
    \end{table*}
    
The last approach in designing the \ac{SNN} accelerators is to target a reconfigurable hardware platform, such as an \ac{FPGA}. The \ac{FPGA} can host different accelerators. This is the reason why many embedded platforms are starting to include them. Second, online reconfigurability allows hardware modification while the system is on. This can be used to add an accelerator after the system has been deployed, remove it once it is no longer required, and modify its functionality. 

Taking \ac{SNN} as an example, the architecture of the network can be modified to target a different set of data if necessary. Finally, partial reconfigurability can add and remove functionalities to a specific accelerator. For example, online learning can be activated and deactivated on request, enabling and disabling the corresponding circuitry and physically adding or removing the required piece of hardware, guaranteeing the optimal architecture for the required application. This is the idea behind \cite{Carpegna:2022aa}, where the authors started to design a tiny hardware accelerator to fit an \ac{FPGA} together with other components employing a \ac{LIF} neuron model. 

The idea is then to have a first degree of reconfigurability, making the accelerator programmable in many aspects, such as weights and thresholds. Then, to add a layer of flexibility by allowing easy modification of the hardwired network hyper-parameters, such as the membrane time constant, the network architecture, the internal bit-widths, etc. Several other works are targeting a more standard but still configurable implementation, such as \cite{minitaur}, \cite{wangQian2017Eepn}, \cite{darwin}, \cite{sixuLi}. Table \ref{tab:FPGAcomparisonTable} compares different accelerators. For more details, see \cite{sixuLi}.

In general, digital accelerators can help a lot in increasing the execution efficiency of \acp{SNN}. CMOS technology is decades old and nowadays widespread, low-cost, and highly optimized. However, the intrinsic behavior of digital devices is very far from that observed in biological components, and the response time and power consumption are still a burden. Augmented silicon photonics platforms can cover most design aspects with more efficient solutions.

\section{Augmented silicon photonics platforms}
\label{sec:photonics}

Integrated photonics is one of the key technologies that has been investigated to build neuromorphic hardware \cite{paquot_optoelectronic_2012,vandoorne_experimental_2014,peng_neuromorphic_2018,feldmann_all-optical_2019}. In particular, \acp{PNN} based on silicon photonics have been extensively investigated for developing lightweight, low-latency, high-speed computing hardware with ultra-low power consumption \cite{shen_deep_2017,nahmias_photonic_2020,feldmann_all-optical_2019,feldmann_parallel_2021}.

Such properties arise from the intrinsic nature of light manipulation and propagation and the capabilities currently integrated photonics platforms can offer. For example, different frequencies of light can be used to encode different data streams separately onto each frequency and then be processed in parallel, thus increasing computing density \cite{peng_neuromorphic_2018}. Such wavelength multiplexing approaches are beneficial for increasing the parallelization degree of architectures. This key feature is used in broadcast and weight protocol where each frequency channel has a specific weight assigned (for each layer) before being summed up together, e.g., by a photodetector \cite{Nahmias2021}. In this approach, ring resonators are key devices enabling the multiplexing (filtering) and weighting of the signals at different frequencies \cite{feldmann_all-optical_2019}.  

Indeed, photonic approaches allow light manipulation (e.g., weights application) while preserving signals propagation at the speed of light throughout the photonic network, thus resulting in ultra-low latencies, limited only by the physical size of the network leading to orders of magnitude lower values compared to electronics implementations  \cite{shen_deep_2017}.

Another essential feature of photonic neuromorphic systems is the possibility of operating with analog complex-valued signals, which is beneficial to leverage non-linearities in neuromorphic hardware such as the electro-optic conversion between complex-valued optical fields into intensities (i.e., photocurrents at the photodetection), but also thanks to connection matrices presenting a more considerable richness in degrees of freedom \cite{vandoorne_experimental_2014,shen_deep_2017,abdalla_minimum_2023}.

Furthermore, \acp{PNN} can operate at much higher speeds than digital accelerators, with their main limitation coming from electro-optic conversion stages, e.g., at the read-out of a \ac{PNN} where photodetectors allow to operate at speeds of hundreds of GHz depending on the technology and responsivity required~\cite{shen_deep_2017}. 

Among the various integrated photonic platforms that have been considered, \ac{SOI} platforms are those that have attracted the most vital interest thanks to the availability of both active and passive components and their reduced footprint compared to platforms with lower refractive indices contrasts, such as \ac{SiNOI}.

\autoref{tab:photonic_vs_electronic} shows a comparison under different metrics for digital accelerators, flash technology, and three different types of \acp{PNN} based on hybrid lasers, co-integrated silicon photonics, and sub-$\lambda$ nanophotonics. For the latter, the device footprint shrinks by at least an order of magnitude due to the robust localization of the optical fields, e.g., in photonic crystal cavities. It is worth noting that such constrained photonic approaches can expect a significant gain in energy consumption and latency. More information on the specific implementations can be found here~\cite{nahmias_photonic_2020}.

\begin{figure}[ht!]
\centerline{\includegraphics[width=0.98\columnwidth]{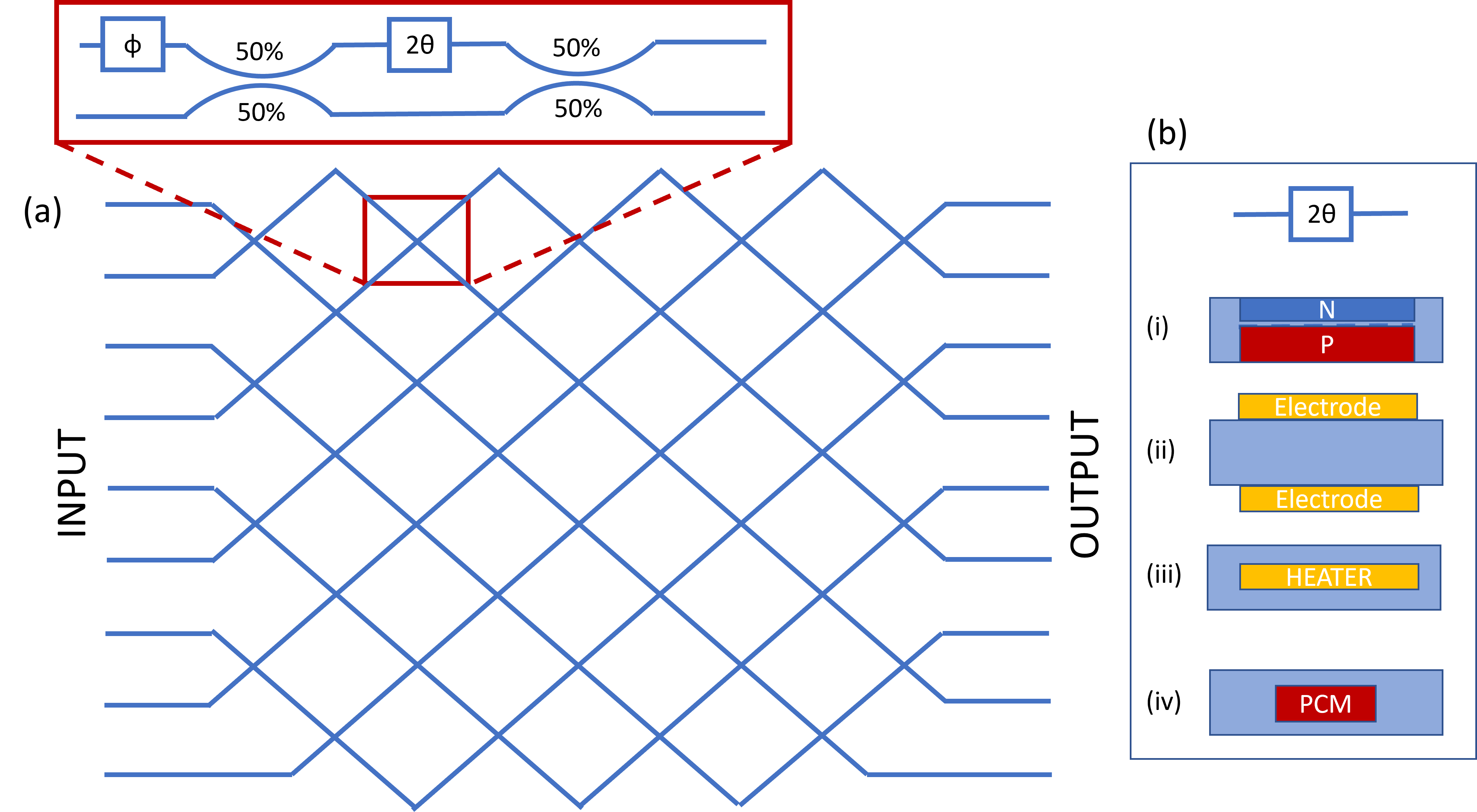}}
\caption{(a) Photonic implementation of a network for 9x9 matrix multiplication based on \acp{MZI}~\cite{clements_optimal_2016}. Inset describes what each crossing consists of, i.e., a phase shifter ($\phi$), then a 50/50 splitter, another phase shifter ($2\theta$), and then a 50/50 combiner. (b) Top-view of phase shifter implementations in waveguides based on (i) thermo-optic effect, (ii) electro-optic effect, (iii) carrier plasma effect, and (iv) \ac{PCM}-induced shift. Blue lines where no crossing is present are optical waveguides.}
\label{fig:phot_neuro}
\end{figure}

\begin{table*}[ht!]

        \caption{Photonic versus electronic approaches comparison table. Latency is the time for a single matrix multiplication operation to compute at the given vector size. Speed is the time between subsequent matrix multiplies~\cite{nahmias_photonic_2020}.}

        \label{tab:photonic_vs_electronic}

        \centering

        \begin{tabular}{|c|c|c|c|c|c|c|}

            \hline

            \textbf{Technology}             & 
            Google TPU  &
            Flash &
            Hybrid laser \ac{NN}			&
            Co-Integrated Si \ac{NN} 		&
            Sub-$\lambda$ nanophotonics           \\

			& 
             \cite{noauthor_-datacenter_nodate} &
            (Analog) \cite{mahmoodi_ultra-low_2018}   &
            \cite{nahmias_leaky_2013}			&
            \cite{tait_silicon_2019}		&
            \cite{nahmias_photonic_2020}             \\

            \hline

            \textbf{Energy/MAC [fJ]}	& 
            430				&
            7  			&
            220			&
            2.7				&
            0.03 			\\

            \hline

            \textbf{Comp. density [TMACs/s/mm$^2$]}		&
            0.58			&
            18		&
            4.5		&
            50		&
            5000		\\

            \hline

            \textbf{Vector size}	&
            256		&
            100			&
            56			&
            148        &
            300       \\

            \hline

            \textbf{Precision [bits]}		&
            8				&
            5				&
            5.1+				&
            5.1+				&
            5.1+				\\

            \hline

            \textbf{Latency/speed [ns]}		&
            2000/1.42			&
            15 			&
            $<$ 0.1   			&
            $<$ 0.1	    		&
            $<$ 0.05			\\

            \hline

        \end{tabular}
        
    \end{table*}

In particular, \ac{SOI} platforms can provide high-speed modulators, e.g., in \ac{MZI} or \ac{RR} configurations, as well as high-speed broadband photodetectors ($>$ 50 GHz) and low propagation losses ($<$ 3 dB/cm) \cite{margalit_perspective_2021}. More specifically, \ac{MZI} devices allow to modulate light and change its amplitude (and phase), therefore providing a practical way to implement \ac{ANN} weights \cite{shen_deep_2017}. They can be arranged in meshes in precise ways, e.g., matrix multiplications as shown in Fig.~\ref{fig:phot_neuro}(a) \cite{clements_optimal_2016}. The traditionally used physical mechanisms for modulation are either (i) thermo-optic, (ii) electro-optic, (iii) carrier plasma effect \cite{bogaerts_silicon_2012} as schematically shown in Fig.~\ref{fig:phot_neuro}(b). For the thermo-optic approach, one of the interferometer arms is heated up by a micro-heater, resulting in a change in the refractive index of the arm. In the second case, electrodes can establish an electric field (for electro-optic materials - not present in native Si platforms) that modifies the refractive index. In the third case, a p-n/p-i-n junction is used where carriers concentration in the depletion region is modified by an applied voltage, thus changing the refractive index by the plasma carrier effect \cite{bogaerts_silicon_2012}. All these approaches are of interest (depending on the platform available) and have enabled vector-matrix multiplication and \acp{PNN} for classification tasks \cite{shen_deep_2017}. 

However, in silicon platforms, one of the main limitations during the inference process is the need to dissipate energy to keep the values of the weights. This energy can account for up to 10 mW with a total $\pi$ phase shift per \ac{MZI}. Such a shift allows routing full signals from one output to the other of an \ac{MZI} with two output ports (see inset of  Fig.~\ref{fig:phot_neuro})(a) starting from a signal only coming from the former port.

In \cite{shen_deep_2017}, the authors used a specific arrangement of mesh devices to perform matrix multiplication through singular value decomposition. This mapping is precisely defined between the elements of the matrix and the phase shifters \acp{MZI} in the mesh. Unlike conventional electronics like CPUs and GPUs, the energy consumed per calculation (\ac{FLOP}) reduces to zero as the size of the mesh grows larger (assuming non-volatile weights). In contrast, conventional electronics require a fixed amount of energy per \ac{FLOP}, and the overall energy consumption scales quadratically with the size of the problem, i.e., $N^2$ instead of just $N$ for \acp{PNN} with $N$ representing the mesh size. The authors in~\cite{shen_deep_2017} also used this \acp{MZI} as weights to build a \ac{FFNN} that could analyze vowels, achieving a simulated correctness of 90\%, comparable to that of a digital computer which would reach 91.7\%.

One solution to avoid constant power dissipation is setting the weights using waveguides integrating non-volatile materials such as \acp{PCM}, e.g., above the waveguide as in (iv) approach in Fig.~\ref{fig:phot_neuro})(b). Their response (change in the degree of crystalline to amorphous ratio) can be set using electrical or optical pulses \cite{rios_controlled_2018}. Such materials are exciting also to implement \ac{STDP} thanks to their very rapid response time (sub-ns) to stimuli, thus allowing to build of optical plastic synapses and spiking architectures based on ring resonators \cite{feldmann_all-optical_2019}.

The recently started Horizon Europe NEUROPULS project investigates a series of approaches that leverage silicon photonics platforms with the addition of \acp{PCM} and III-V materials for building more efficient neuromorphic hardware.

An approach to improving CNN performance is to use an accelerator designed for matrix-vector products, such as a mesh of modulators programmed to perform a specific matrix multiplication. Singular value decomposition can factor the matrix, resulting in more efficient computation on the accelerator. This approach can also handle other matrix operations, including convolutions, and optimize hardware architecture for specific tasks, leading to significant training and inference times speed-ups \cite{shen_deep_2017}.

Although optical components can benefit matrix operations, their size can limit the size of matrices that can be implemented. To overcome this limitation, alternative approaches can be explored. One possible solution is to employ pruning techniques that remove unnecessary connections and weights from the network, thereby reducing the overall size of the matrix. This makes it possible to implement larger systems using the available optical components.

Another option is to use block matrix decompositions, which involve dividing the matrix into smaller blocks that can be processed separately. This approach can enable the implementation of larger matrices using a smaller number of optical components. The smaller blocks can be computed independently and combined for the final result. This technique can also be combined with other optimization strategies, such as quantization and compression, further to reduce the size and complexity of the system.

The tensor-train approach proposed in \cite{xiaoLargescaleEnergyefficientTensorized2021} will be explored. This approach represents the matrix as a product of low-rank tensors, which can be processed more efficiently using optical components. The goal is to develop a scalable optical architecture capable of handling larger matrices and more complex neural networks using the abovementioned approaches. This will facilitate faster and more efficient training of neural networks using photonics.

The research will also explore \ac{RNN} applications, including fully trainable \acp{RNN} as well as \acp{RC}. We will investigate the potential applications of \ac{RC} in photonics for various applications, including nonlinear dispersion compensation of telecom signals, as demonstrated in \cite{sackesyn_experimental_2021}. The research aims to develop new techniques and approaches for utilizing photonics in \ac{ML} and other fields, which could lead to significant advancements in the performance and efficiency of these systems.

To implement non-volatile optical weights, the proposed architecture will incorporate \acp{PCM}. Previous studies, such as \cite{feldmann_all-optical_2019} and \cite{Miscuglio:2020aa}, have explored these materials and shown a strong potential for neuromorphic systems. Incorporating non-volatile weights can significantly reduce power consumption compared to volatile weights, which require continuous driving or periodic refreshing. Including these materials in the proposed architecture will be crucial in developing low-power and high-performance systems for machine learning and other applications.

However, in addition to their non-volatility, \acp{PCM} have another advantage: their nonlinear dynamics. E.g., by exciting the material with pulses rather than continuous-wave excitation, the nonlinear behavior of the material enables other computing paradigms, such as \acp{SNN}. In such networks, the neurons communicate using brief pulses or spikes rather than continuously varying signals. This opens up the implementation of energy-efficient and highly parallel neural networks. To generate the spikes injected into the system, Advanced high-extinction ratio (ER $>$ 8 dB) Q-switched spiking lasers can be used, which will be monolithically integrated into III-V materials on the same platform \cite{10.3389/fphy.2022.1017714}. These hybrid III-V-on-Si spiking lasers are a scalable and cost-effective alternative to previous Q-switched lasers made purely from III-V materials.

III-V-on-Si spiking lasers will generate highly precise and controlled optical spikes, essential for many photonics and \ac{ML} applications. These lasers offer several advantages, including high extinction ratios, low power consumption, and compatibility with standard silicon processing techniques.

\section{Reliability studies and concerns}
\label{sec:reliability}

\acp{ANN} have an intrinsic error tolerance from an algorithmic point of view thanks to their redundant nature. However, hardware designs to deploy such algorithms must be analyzed to assess the impact of hardware restrictions or faulty manifestation on the network functional's behavior. 

Due to manufacturing issues, hardware faults can occur randomly, provoked by neuron and synapse defects and imprecisions. Still, they can also be malicious, introduced by different kinds of attacks (i.e., laser beams fault injection or row hammer attack) \cite{8203770}, \cite{breier2018deeplaser}. The authors of \cite{breier2018deeplaser} have analyzed the misclassification rate of \ac{MLP} based deep neural networks face to models derived from physical phenomena. Faulty-neuron behaviors have been injected randomly or deterministically, with injection scenarios considering single and multiple faulty neurons per layer. This is usually done during the function activation timeframe, which can be hundreds to thousands of cycles. Results indicate that in some cases, even a relatively small number of faulty neurons ($\approx$ 10\%) can lead to a high risk of misclassification ($\approx$ 62\%). As seen in \autoref{sec:background}, different activation functions have been studied for \acp{DNN}. They show that for a higher miss-classification rate ($>$50\%), at least half of the neurons in a given hidden layer should be faulty, which is the case for sigmoid and $\tanh$ activation functions. In the case of \ac{ReLU}, at least 3/4 of the neurons should be faulty to achieve the same miss-classification rate. 

Extensive work has been dedicated in the last years to studying and evaluating AI hardware accelerators' errors and fault tolerance. An overview of fault tolerance techniques for feedforward neural networks is presented in \cite{8013784}. In this paper, the authors review fault types, models, and measures used to evaluate performance and provide a taxonomy of the main techniques to enhance the intrinsic properties of some neural models based on the principles and mechanisms they exploit to achieve fault tolerance passively. 

In \cite{8645906}, the authors present a study of the fault characterization and mitigation of \ac{RTL} model of NN accelerators by characterizing the vulnerability of NNs to application-level specifications, network topology, and activation functions, as well as architectural level specifications. In \cite{9926241}, authors present an experimental evaluation of the resilience of \ac{DNN} systems (i.e., \ac{DNN} software running on specialized accelerators) under Soft errors caused by high-energy particles. 

An empirical study of \acp{DNN} resilience can be found in \cite{8465834}, where a fault injection framework named Ares, which can deal with fully connected and \ac{CNN}-based \ac{DNN} accelerators, is presented. It uses hardware fault models related to technology and environment variability, single event faults transients in memory elements, and algorithmic level faults models such as faults occurring in weights, activation, and hidden states. Fault injection is performed static, offline, before the inference process, and dynamically during inference execution.

The analysis of \acp{SNN} fault tolerance and reliability is a relatively newer field of research since their hardware implementations are much more recent than classical \acp{DNN}. Consensus has yet to be reached on the main applications of these networks. Moreover, the variety of signal-to-spike coding and training algorithms brings a specific heterogeneity in their characteristic which should be accounted for when their fault tolerance and reliability is discussed. One of the first works in this field, \cite{8053727}, estimates the accuracy of a \ac{FC} \ac{SNN}, capable of \ac{STDP} learning designed with spintronic devices under the effect of process variability. Both the neuron and synapse behavior are strongly affected by process variability, and the accuracy drops by approximately 10\% when assuming moderate variability compared to the ideal case. 

Another interesting study is presented in \cite{8758653}, where a taxonomy of faults was defined for spiking neural networks. The accuracy of a hardware-implemented spiking neural network designed to perform \ac{STDP} online training was analyzed under the assumption that (i) both learning and inference were performed on faulty hardware, (ii) only inference was performed on faulty hardware. This paper shows that performing learning directly on faulty hardware reduces the impact of faults on the network accuracy by an average of 15\% and, in extreme cases, can reach up to 30\%. Moreover, in \cite{8875270}, faults affecting the signal-to-spike conversion layer have the strongest impact on the network accuracy, with the synaptic stuck-at faults coming to a very close second. These faults strongly affect the learning process, which only exacerbates during inference.

On the other hand, faults, like delayed synapse activation or stuck lateral inhibition, have a marginal effect. The study presented in \cite{9774711} takes a different approach. The fault tolerance study is performed on an \ac{SNN} inference accelerator, a multi-layer \ac{SNN} with supervised off-line learning, designed in VHDL and implemented on an \ac{FPGA}. The fault injection experiments identify the parts of the design that need to be protected against faults and the inherently fault-tolerant parts. They have shown that the behavior under faults of the chosen type of \ac{SNN} is similar to the behavior under faults of an \ac{ANN} in that faults injected in the most significant bits of the synaptic values affect have a more substantial effect on network accuracy that when the faults are injected in the less critical bits, and also faults affecting the last layer (where the classification is performed) are more relevant than faults affecting the first layer (where there is higher computing redundancy). In addition, the authors have shown that their proposed \ac{SNN} implementation is much more sensitive to faults injected in the routing of signals than in the synaptic weight or neuron computation.

\begin{figure}[htb]
\centerline{\includegraphics[width=0.47\textwidth]{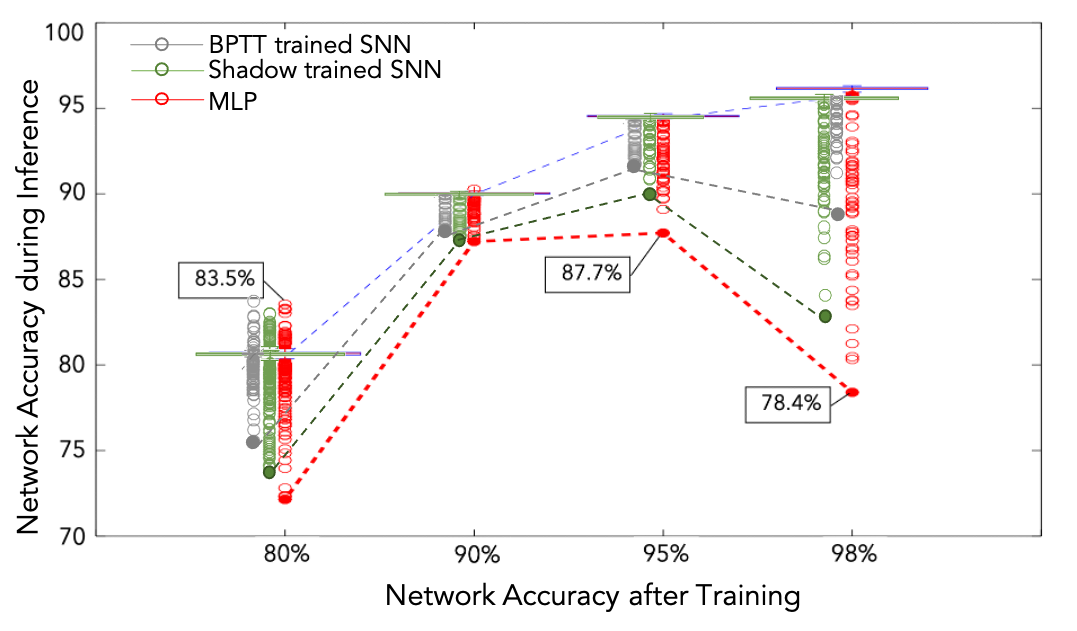}}
\caption{Network accuracy under synaptic fault injection: a comparison between \ac{MLP} and \ac{SNN}. The synaptic weight precision is 8 bits, and $10^{-4}$ of the synaptic bits are considered faulty. }
\label{SNNvsMLP}
    \vspace{-0.5cm}
\end{figure}

\begin{figure}[htb]
\centerline{\includegraphics[width=0.47\textwidth]{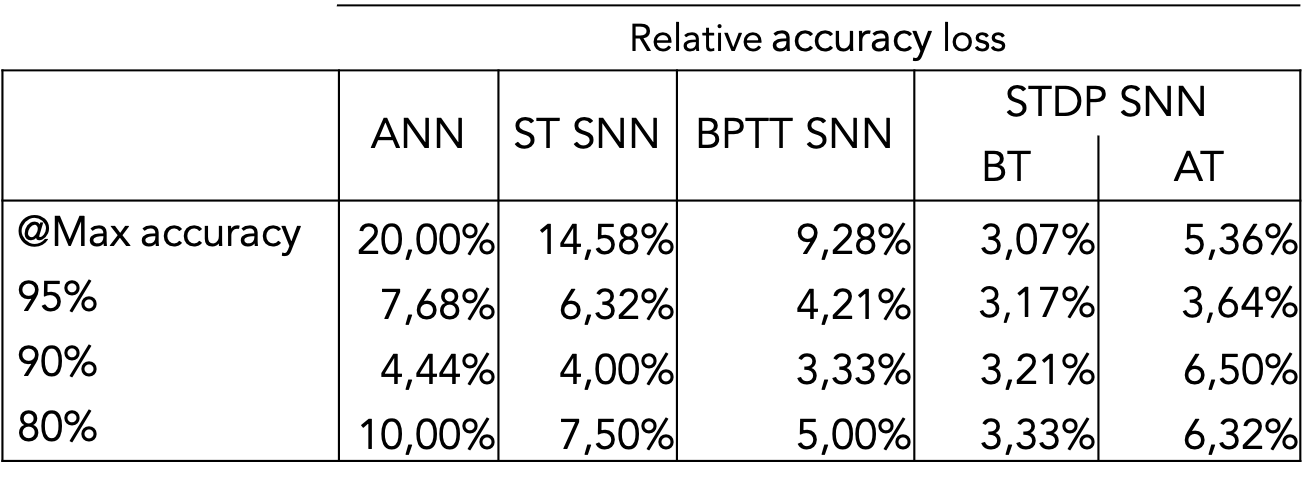}}
\caption{Maximum network accuracy degradation for different training scenarios under synaptic fault injection: a comparison between \ac{MLP} and \ac{SNN}. The synaptic weight precision is 8 bits, and $10^{-4}$ of the synaptic bits are considered faulty. AT = after training, BT = before training}
\label{SNNvsMLP2}
    \vspace{-0.5cm}
\end{figure}

Several studies have compared \acp{SNN} and their \acp{ANN} counterparts, mainly focusing on performance or power consumption rather than their relative fault tolerance. Therefore, a study was conducted to assess the fault resilience of both network types, assuming different quantizations of weights and neural computation and various training scenarios for the \ac{SNN}. The study compared the fault tolerance of an \ac{MLP} and an \ac{SNN} with the same topology and bit precision. A 784 X 100 X 10 network was implemented to solve the MNIST classification problem \cite{LeCun1998}. The \ac{MLP} uses a sigmoid activation function for all layers and has a base accuracy of 98\% after 20 training epochs. The \ac{SNN} used rate coding for signal-to-spike conversion and several training algorithms: (i) \ac{ST} (with a base accuracy of 96\%), (ii) \ac{BPTT} (with a base accuracy of 97\%), (iii) \ac{STDP} (with a base accuracy of 87\%). The precision of synaptic weights was 16, 8, and 4 bits. Figure \ref{SNNvsMLP} illustrates selected results of the analysis, showing that the \acp{SNN} are more fault-tolerant than the \ac{MLP}, and the \ac{SNN} trained using \ac{BPTT} is the most resilient of the three. This study also demonstrates how the \ac{SNN} is trained in its fault resilience, with online training being more resilient than offline training. The last two columns in \autoref{SNNvsMLP2} correspond to the SNN being trained by \ac{STDP}, and the network accuracy degradation is computed when the presence of faults is assumed only at inference (AT) or during both training and inference (BT). These results show that the fault tolerance of the \ac{SNN} depends on the training mechanism and that an \ac{SNN} trained online is more resilient to faults than an \ac{SNN} trained offline. The results suggest that \acp{SNN} have the potential to be more resilient to faults than \acp{ANN}, and further comparison between the two is necessary to consolidate these findings.

\section{Conclusion}
\label{sec:conclusion}

In this paper, the fundamentals of \acl{CNN}, \acl{DNN}, and \acl{SNN} were introduced, and their digital counterparts were described using standard CMOS technologies. The discussion then focused on augmented silicon photonics improvements and the reliability aspects of \acl{ANN}. The benefits and drawbacks of each technology were thoroughly analyzed, including its reliability. Our analysis suggests that the choice of technology for \acl{ANN} design should depend on the specific application requirements and design constraints. While CMOS technology offers established fabrication processes and high reliability, it may have limitations in power consumption and scalability. Silicon photonics, on the other hand, provides high power efficiency and scalability, but their reliability is still under ongoing research. We hope this paper will serve as a valuable reference for researchers and practitioners in the field of \acl{ANN} design as they explore the potential of these emerging technologies.

\bibliographystyle{IEEEtran}
\bibliography{biblio}

% Generated by IEEEtran.bst, version: 1.14 (2015/08/26)
\begin{thebibliography}{10}
\providecommand{\url}[1]{#1}
\csname url@samestyle\endcsname
\providecommand{\newblock}{\relax}
\providecommand{\bibinfo}[2]{#2}
\providecommand{\BIBentrySTDinterwordspacing}{\spaceskip=0pt\relax}
\providecommand{\BIBentryALTinterwordstretchfactor}{4}
\providecommand{\BIBentryALTinterwordspacing}{\spaceskip=\fontdimen2\font plus
\BIBentryALTinterwordstretchfactor\fontdimen3\font minus
  \fontdimen4\font\relax}
\providecommand{\BIBforeignlanguage}[2]{{%
\expandafter\ifx\csname l@#1\endcsname\relax
\typeout{** WARNING: IEEEtran.bst: No hyphenation pattern has been}%
\typeout{** loaded for the language `#1'. Using the pattern for}%
\typeout{** the default language instead.}%
\else
\language=\csname l@#1\endcsname
\fi
#2}}
\providecommand{\BIBdecl}{\relax}
\BIBdecl

\bibitem{Marchisio:2019aa}
A.~Marchisio \emph{et~al.}, ``Deep learning for edge computing: Current trends,
  cross-layer optimizations, and open research challenges,'' in \emph{2019 IEEE
  Computer Society Annual Symposium on VLSI (ISVLSI)}.\hskip 1em plus 0.5em
  minus 0.4em\relax IEEE, 2019, pp. 553--559.

\bibitem{Shafique:2018aa}
M.~Shafique \emph{et~al.}, ``An overview of next-generation architectures for
  machine learning: Roadmap, opportunities and challenges in the iot era,'' in
  \emph{2018 Design, Automation \& Test in Europe Conference \& Exhibition
  (DATE)}.\hskip 1em plus 0.5em minus 0.4em\relax IEEE, 2018, pp. 827--832.

\bibitem{Dutto:2021aa}
S.~Dutto \emph{et~al.}, ``Exploring deep learning for in-field fault detection
  in microprocessors,'' in \emph{2021 Design, Automation \& Test in Europe
  Conference \& Exhibition (DATE)}, 2021, pp. 1456--1459.

\bibitem{Sedjelmaci:2016aa}
H.~Sedjelmaci \emph{et~al.}, ``A lightweight anomaly detection technique for
  low-resource iot devices: A game-theoretic methodology,'' in \emph{2016 IEEE
  International Conference on Communications (ICC)}, 2016, pp. 1--6.

\bibitem{zhou2018brief}
C.-W. Zhou \emph{et~al.}, ``A brief survey on deep neural networks,''
  \emph{arXiv preprint arXiv:1802.08882}, 2018.

\bibitem{LeCun1998}
Y.~LeCun \emph{et~al.}, ``Gradient-based learning applied to document
  recognition,'' in \emph{Proceedings of the IEEE}, vol.~86, no.~11.\hskip 1em
  plus 0.5em minus 0.4em\relax IEEE, 1998, pp. 2278--2324.

\bibitem{goodfellow2016deep}
I.~Goodfellow \emph{et~al.}, ``Deep learning,'' in \emph{Deep learning}.\hskip
  1em plus 0.5em minus 0.4em\relax MIT Press, 2016, pp. 1--800.

\bibitem{schuman_opportunities_2022}
\BIBentryALTinterwordspacing
C.~D. Schuman \emph{et~al.}, ``\BIBforeignlanguage{en}{Opportunities for
  neuromorphic computing algorithms and applications},''
  \emph{\BIBforeignlanguage{en}{Nat Comput Sci}}, vol.~2, no.~1, pp. 10--19,
  Jan. 2022, number: 1 Publisher: Nature Publishing Group. [Online]. Available:
  \url{https://www.nature.com/articles/s43588-021-00184-y}
\BIBentrySTDinterwordspacing

\bibitem{ThirdGen}
W.~Maass, ``Networks of spiking neurons: The third generation of neural network
  models,'' \emph{Neural networks}, vol.~10, no.~9, 1997.

\bibitem{Heildelberg}
B.~Cramer \emph{et~al.}, ``The heidelberg spiking data sets for the systematic
  evaluation of spiking neural networks,'' \emph{IEEE Transactions on Neural
  Networks and Learning Systems}, vol.~33, no.~7, pp. 2744--2757, 2022.

\bibitem{EcgSnn}
A.~Amirshahi and M.~Hashemi, ``Ecg classification algorithm based on stdp and
  r-stdp neural networks for real-time monitoring on ultra low-power personal
  wearable devices,'' \emph{IEEE Transactions on Biomedical Circuits and
  Systems}, vol.~13, no.~6, pp. 1483--1493, 2019.

\bibitem{Vallero:2019aa}
A.~Vallero \emph{et~al.}, ``Syra: Early system reliability analysis for
  cross-layer soft errors resilience in memory arrays of microprocessor
  systems,'' \emph{IEEE Transactions on Computers}, vol.~68, no.~5, pp.
  765--783, 2019.

\bibitem{Ruospo:2023aa}
A.~Ruospo \emph{et~al.}, ``A survey on deep learning resilience assessment
  methodologies,'' \emph{Computer}, vol.~56, no.~2, pp. 57--66, 2023.

\bibitem{james_historical_2017}
\BIBentryALTinterwordspacing
C.~D. James \emph{et~al.}, ``\BIBforeignlanguage{en}{A historical survey of
  algorithms and hardware architectures for neural-inspired and neuromorphic
  computing applications},'' \emph{\BIBforeignlanguage{en}{Biologically
  Inspired Cognitive Architectures}}, vol.~19, pp. 49--64, Jan. 2017. [Online].
  Available:
  \url{https://www.sciencedirect.com/science/article/pii/S2212683X16300561}
\BIBentrySTDinterwordspacing

\bibitem{Krizhevsky2012}
\BIBentryALTinterwordspacing
A.~Krizhevsky \emph{et~al.}, ``Imagenet classification with deep convolutional
  neural networks,'' \emph{Commun. ACM}, vol.~60, no.~6, pp. 84--90, may 2017.
  [Online]. Available: \url{https://doi.org/10.1145/3065386}
\BIBentrySTDinterwordspacing

\bibitem{rumelhart1986learning}
D.~E. Rumelhart \emph{et~al.}, ``Learning representations by back-propagating
  errors,'' \emph{Nature}, vol. 323, no. 6088, pp. 533--536, 1986.

\bibitem{Hinton:2006aa}
\BIBentryALTinterwordspacing
G.~E. Hinton and R.~R. Salakhutdinov, ``Reducing the dimensionality of data
  with neural networks,'' \emph{Science}, vol. 313, no. 5786, pp. 504--507,
  2006. [Online]. Available:
  \url{https://www.science.org/doi/abs/10.1126/science.1127647}
\BIBentrySTDinterwordspacing

\bibitem{TrueNorth}
F.~Akopyan \emph{et~al.}, ``Truenorth: Design and tool flow of a 65 mw 1
  million neuron programmable neurosynaptic chip,'' \emph{IEEE Transactions on
  Computer-Aided Design of Integrated Circuits and Systems}, vol.~34, no.~10,
  2015.

\bibitem{DRAGHICI2002395}
\BIBentryALTinterwordspacing
S.~Draghici, ``On the capabilities of neural networks using limited precision
  weights,'' \emph{Neural Networks}, vol.~15, no.~3, pp. 395--414, 2002.
  [Online]. Available:
  \url{https://www.sciencedirect.com/science/article/pii/S0893608002000321}
\BIBentrySTDinterwordspacing

\bibitem{doi:10.1126/science.1254642}
\BIBentryALTinterwordspacing
P.~A. Merolla \emph{et~al.}, ``A million spiking-neuron integrated circuit with
  a scalable communication network and interface,'' \emph{Science}, vol. 345,
  no. 6197, pp. 668--673, 2014. [Online]. Available:
  \url{https://www.science.org/doi/abs/10.1126/science.1254642}
\BIBentrySTDinterwordspacing

\bibitem{HodgkinHuxley}
A.~L. Hodgkin and A.~F. Huxley, ``A quantitative description of membrane
  current and its application to conduction and excitation in nerve,''
  \emph{The Journal of Physiology}, vol. 117, no.~4, pp. 500--544, 1952.

\bibitem{Izhikevich}
E.~Izhikevich, ``Simple model of spiking neurons,'' \emph{IEEE Transactions on
  Neural Networks}, vol.~14, no.~6, pp. 1569--1572, 2003.

\bibitem{Glif}
C.~Teeter \emph{et~al.}, ``Generalized leaky integrate-and-fire models classify
  multiple neuron types,'' \emph{Nature Communications}, vol.~9, no.~1, p. 709,
  2018.

\bibitem{IfReview}
A.~N. Burkitt, ``A review of the integrate-and-fire neuron model: Ii.
  inhomogeneous synaptic input and network properties,'' \emph{Biological
  Cybernetics}, vol.~95, no.~2, pp. 97--112, 2006.

\bibitem{Carpegna:2022aa}
A.~Carpegna \emph{et~al.}, ``Spiker: an fpga-optimized hardware accelerator for
  spiking neural networks,'' in \emph{2022 IEEE Computer Society Annual
  Symposium on VLSI (ISVLSI)}, 2022, pp. 14--19.

\bibitem{SnnReservoir}
F.~Corradi and G.~Indiveri, ``A neuromorphic event-based neural recording
  system for smart brain-machine-interfaces,'' \emph{IEEE Transactions on
  Biomedical Circuits and Systems}, vol.~9, no.~5, pp. 699--709, 2015.

\bibitem{vandersandeAdvancesPhotonicReservoir2017}
\BIBentryALTinterwordspacing
G.~V. der Sande \emph{et~al.}, ``Advances in photonic reservoir computing,''
  \emph{Nanophotonics}, vol.~6, no.~3, pp. 561--576, 2017. [Online]. Available:
  \url{https://doi.org/10.1515/nanoph-2016-0132}
\BIBentrySTDinterwordspacing

\bibitem{RateCoding}
D.~J. Enoka~RM, ``Rate coding and the control of muscle force,'' \emph{Cold
  Spring Harb Perspect Med}, 2017.

\bibitem{TemporalCoding}
B.~Petro \emph{et~al.}, ``Selection and optimization of temporal spike encoding
  methods for spiking neural networks,'' \emph{IEEE Transactions on Neural
  Networks and Learning Systems}, vol.~31, no.~2, pp. 358--370, 2020.

\bibitem{PopulationCoding}
Z.~Pan \emph{et~al.}, ``Neural population coding for effective temporal
  classification,'' in \emph{2019 International Joint Conference on Neural
  Networks (IJCNN)}, 2019, pp. 1--8.

\bibitem{SGD}
L.~Bottou, ``Large-scale machine learning with stochastic gradient descent,''
  in \emph{Proceedings of COMPSTAT'2010}, Y.~Lechevallier and G.~Saporta,
  Eds.\hskip 1em plus 0.5em minus 0.4em\relax Heidelberg: Physica-Verlag HD,
  2010, pp. 177--186.

\bibitem{AdaGrad}
J.~Duchi \emph{et~al.}, ``Adaptive subgradient methods for online learning and
  stochastic optimization,'' \emph{J. Mach. Learn. Res.}, vol.~12, no. null,
  pp. 2121--2159, jul 2011.

\bibitem{Dropout}
N.~Srivastava \emph{et~al.}, ``Dropout: A simple way to prevent neural networks
  from overfitting,'' \emph{J. Mach. Learn. Res.}, vol.~15, no.~1, pp.
  1929--1958, jan 2014.

\bibitem{Takahashi:2018aa}
\BIBentryALTinterwordspacing
R.~Takahashi \emph{et~al.}, ``Ricap: Random image cropping and patching data
  augmentation for deep cnns,'' in \emph{Proceedings of The 10th Asian
  Conference on Machine Learning}, ser. Proceedings of Machine Learning
  Research, J.~Zhu and I.~Takeuchi, Eds., vol.~95.\hskip 1em plus 0.5em minus
  0.4em\relax PMLR, 14--16 Nov 2018, pp. 786--798. [Online]. Available:
  \url{https://proceedings.mlr.press/v95/takahashi18a.html}
\BIBentrySTDinterwordspacing

\bibitem{Fischer:2012aa}
A.~Fischer and C.~Igel, ``An introduction to restricted boltzmann machines,''
  in \emph{Progress in Pattern Recognition, Image Analysis, Computer Vision,
  and Applications}, L.~Alvarez \emph{et~al.}, Eds.\hskip 1em plus 0.5em minus
  0.4em\relax Berlin, Heidelberg: Springer Berlin Heidelberg, 2012, pp. 14--36.

\bibitem{TransferLearning}
S.~J. Pan and Q.~Yang, ``A survey on transfer learning,'' \emph{IEEE
  Transactions on Knowledge and Data Engineering}, vol.~22, pp. 1345--1359,
  2010.

\bibitem{RL}
V.~Mnih \emph{et~al.}, ``Playing atari with deep reinforcement learning,''
  2013.

\bibitem{BiPooStdp}
P.~M. Bi~GQ, ``Synaptic modifications in cultured hippocampal neurons:
  dependence on spike timing, synaptic strength, and postsynaptic cell type,''
  \emph{The Journal of Neuroscience}, vol.~18, no.~24, 1998.

\bibitem{SnnTorch}
J.~K. Eshraghian \emph{et~al.}, ``Training spiking neural networks using
  lessons from deep learning,'' \emph{arXiv preprint arXiv:2109.12894}, 2021.

\bibitem{Slayer}
\BIBentryALTinterwordspacing
S.~B. Shrestha and G.~Orchard, ``{SLAYER}: Spike layer error reassignment in
  time,'' in \emph{Advances in Neural Information Processing Systems 31},
  S.~Bengio \emph{et~al.}, Eds.\hskip 1em plus 0.5em minus 0.4em\relax Curran
  Associates, Inc., 2018, pp. 1419--1428. [Online]. Available:
  \url{http://papers.nips.cc/paper/7415-slayer-spike-layer-error-reassignment-in-time.pdf}
\BIBentrySTDinterwordspacing

\bibitem{Carlsim}
T.-S. Chou \emph{et~al.}, ``Carlsim 4: An open source library for large scale,
  biologically detailed spiking neural network simulation using heterogeneous
  clusters,'' in \emph{2018 International Joint Conference on Neural Networks
  (IJCNN)}, 2018, pp. 1--8.

\bibitem{SpiNNaker}
S.~B. Furber \emph{et~al.}, ``The spinnaker project,'' \emph{Proceedings of the
  IEEE}, vol. 102, no.~5, 2014.

\bibitem{Loihi}
M.~Davies \emph{et~al.}, ``Loihi: A neuromorphic manycore processor with
  on-chip learning,'' \emph{IEEE Micro}, vol.~38, no.~1, 2018.

\bibitem{Tianjic}
J.~Pei \emph{et~al.}, ``Towards artificial general intelligence with hybrid
  tianjic chip architecture,'' \emph{Nature}, vol. 572, no. 7767, 2019.

\bibitem{Odin}
C.~Frenkel \emph{et~al.}, ``A 0.086-mm$^2$ 12.7-pj/sop 64k-synapse 256-neuron
  online-learning digital spiking neuromorphic processor in 28-nm cmos,''
  \emph{IEEE Transactions on Biomedical Circuits and Systems}, vol.~13, no.~1,
  2019.

\bibitem{SNN_HW_review}
A.~Basu \emph{et~al.}, ``Spiking neural network integrated circuits: A review
  of trends and future directions,'' in \emph{2022 IEEE Custom Integrated
  Circuits Conference (CICC)}, 2022.

\bibitem{sixuLi}
S.~Li \emph{et~al.}, ``A fast and energy-efficient snn processor with adaptive
  clock/event-driven computation scheme and online learning,'' \emph{IEEE
  transactions on circuits and systems. I, Regular papers}, vol.~68, no.~4, pp.
  1543--1552, 2021.

\bibitem{minitaur}
D.~Neil and S.-C. Liu, ``Minitaur, an event-driven fpga-based spiking network
  accelerator,'' \emph{IEEE transactions on very large scale integration (VLSI)
  systems}, vol.~22, no.~12, pp. 2621--2628, 2014.

\bibitem{wangQian2017Eepn}
Q.~Wang \emph{et~al.}, ``Energy efficient parallel neuromorphic architectures
  with approximate arithmetic on fpga,'' \emph{Neurocomputing (Amsterdam)},
  vol. 221, pp. 146--158, 2017.

\bibitem{darwin}
D.~Ma \emph{et~al.}, ``Darwin: A neuromorphic hardware co-processor based on
  spiking neural networks,'' \emph{Journal of systems architecture}, vol.~77,
  pp. 43--51, 2017.

\bibitem{paquot_optoelectronic_2012}
\BIBentryALTinterwordspacing
Y.~Paquot \emph{et~al.}, ``Optoelectronic {Reservoir} {Computing},''
  \emph{Scientific Reports}, vol.~2, no.~1, p. 287, Feb. 2012. [Online].
  Available: \url{https://doi.org/10.1038/srep00287}
\BIBentrySTDinterwordspacing

\bibitem{vandoorne_experimental_2014}
\BIBentryALTinterwordspacing
K.~Vandoorne \emph{et~al.}, ``\BIBforeignlanguage{en}{Experimental
  demonstration of reservoir computing on a silicon photonics chip},''
  \emph{\BIBforeignlanguage{en}{Nat Commun}}, vol.~5, no.~1, p. 3541, Mar.
  2014, number: 1 Publisher: Nature Publishing Group. [Online]. Available:
  \url{https://www.nature.com/articles/ncomms4541}
\BIBentrySTDinterwordspacing

\bibitem{peng_neuromorphic_2018}
\BIBentryALTinterwordspacing
H.-T. Peng \emph{et~al.}, ``\BIBforeignlanguage{en}{Neuromorphic {Photonic}
  {Integrated} {Circuits}},'' \emph{\BIBforeignlanguage{en}{IEEE J. Select.
  Topics Quantum Electron.}}, vol.~24, no.~6, pp. 1--15, Nov. 2018. [Online].
  Available: \url{https://ieeexplore.ieee.org/document/8364605/}
\BIBentrySTDinterwordspacing

\bibitem{feldmann_all-optical_2019}
\BIBentryALTinterwordspacing
J.~Feldmann \emph{et~al.}, ``\BIBforeignlanguage{en}{All-optical spiking
  neurosynaptic networks with self-learning capabilities},''
  \emph{\BIBforeignlanguage{en}{Nature}}, vol. 569, no. 7755, pp. 208--214, May
  2019. [Online]. Available:
  \url{http://www.nature.com/articles/s41586-019-1157-8}
\BIBentrySTDinterwordspacing

\bibitem{shen_deep_2017}
\BIBentryALTinterwordspacing
Y.~Shen \emph{et~al.}, ``\BIBforeignlanguage{en}{Deep learning with coherent
  nanophotonic circuits},'' \emph{\BIBforeignlanguage{en}{Nature Photon}},
  vol.~11, no.~7, pp. 441--446, Jul. 2017. [Online]. Available:
  \url{http://www.nature.com/articles/nphoton.2017.93}
\BIBentrySTDinterwordspacing

\bibitem{nahmias_photonic_2020}
\BIBentryALTinterwordspacing
M.~A. Nahmias \emph{et~al.}, ``\BIBforeignlanguage{en}{Photonic
  {Multiply}-{Accumulate} {Operations} for {Neural} {Networks}},''
  \emph{\BIBforeignlanguage{en}{IEEE J. Select. Topics Quantum Electron.}},
  vol.~26, no.~1, pp. 1--18, Jan. 2020. [Online]. Available:
  \url{https://ieeexplore.ieee.org/document/8844098/}
\BIBentrySTDinterwordspacing

\bibitem{feldmann_parallel_2021}
\BIBentryALTinterwordspacing
J.~Feldmann \emph{et~al.}, ``\BIBforeignlanguage{en}{Parallel convolutional
  processing using an integrated photonic tensor core},''
  \emph{\BIBforeignlanguage{en}{Nature}}, vol. 589, no. 7840, pp. 52--58, Jan.
  2021. [Online]. Available:
  \url{http://www.nature.com/articles/s41586-020-03070-1}
\BIBentrySTDinterwordspacing

\bibitem{Nahmias2021}
Y.~Nahmias and Y.~Loewenstein, ``Neuromodulation of hebbian plasticity:
  relevance and mechanisms,'' \emph{Current Opinion in Neurobiology}, vol.~69,
  pp. 150--159, 2021.

\bibitem{abdalla_minimum_2023}
\BIBentryALTinterwordspacing
M.~Abdalla \emph{et~al.}, ``\BIBforeignlanguage{en}{Minimum complexity
  integrated photonic architecture for delay-based reservoir computing},''
  \emph{\BIBforeignlanguage{en}{Opt. Express}}, vol.~31, no.~7, p. 11610, Mar.
  2023. [Online]. Available:
  \url{https://opg.optica.org/abstract.cfm?URI=oe-31-7-11610}
\BIBentrySTDinterwordspacing

\bibitem{clements_optimal_2016}
\BIBentryALTinterwordspacing
W.~R. Clements \emph{et~al.}, ``\BIBforeignlanguage{en}{Optimal design for
  universal multiport interferometers},''
  \emph{\BIBforeignlanguage{en}{Optica}}, vol.~3, no.~12, p. 1460, Dec. 2016.
  [Online]. Available:
  \url{https://opg.optica.org/abstract.cfm?URI=optica-3-12-1460}
\BIBentrySTDinterwordspacing

\bibitem{noauthor_-datacenter_nodate}
\BIBentryALTinterwordspacing
``In-{Datacenter} {Performance} {Analysis} of a {Tensor} {Processing} {Unit}
  {\textbar} {Proceedings} of the 44th {Annual} {International} {Symposium} on
  {Computer} {Architecture}.'' [Online]. Available:
  \url{https://dl.acm.org/doi/10.1145/3079856.3080246}
\BIBentrySTDinterwordspacing

\bibitem{mahmoodi_ultra-low_2018}
\BIBentryALTinterwordspacing
M.~R. Mahmoodi and D.~Strukov, ``\BIBforeignlanguage{en}{An ultra-low energy
  internally analog, externally digital vector-matrix multiplier based on {NOR}
  flash memory technology},'' in \emph{\BIBforeignlanguage{en}{Proceedings of
  the 55th {Annual} {Design} {Automation} {Conference}}}.\hskip 1em plus 0.5em
  minus 0.4em\relax San Francisco California: ACM, Jun. 2018, pp. 1--6.
  [Online]. Available: \url{https://dl.acm.org/doi/10.1145/3195970.3195989}
\BIBentrySTDinterwordspacing

\bibitem{nahmias_leaky_2013}
M.~A. Nahmias \emph{et~al.}, ``A {Leaky} {Integrate}-and-{Fire} {Laser}
  {Neuron} for {Ultrafast} {Cognitive} {Computing},'' \emph{IEEE Journal of
  Selected Topics in Quantum Electronics}, vol.~19, no.~5, pp. 1--12, Sep.
  2013, conference Name: IEEE Journal of Selected Topics in Quantum
  Electronics.

\bibitem{tait_silicon_2019}
\BIBentryALTinterwordspacing
A.~N. Tait \emph{et~al.}, ``Silicon {Photonic} {Modulator} {Neuron},''
  \emph{Phys. Rev. Appl.}, vol.~11, no.~6, p. 064043, Jun. 2019, publisher:
  American Physical Society. [Online]. Available:
  \url{https://link.aps.org/doi/10.1103/PhysRevApplied.11.064043}
\BIBentrySTDinterwordspacing

\bibitem{margalit_perspective_2021}
\BIBentryALTinterwordspacing
N.~Margalit \emph{et~al.}, ``\BIBforeignlanguage{en}{Perspective on the future
  of silicon photonics and electronics},'' \emph{\BIBforeignlanguage{en}{Appl.
  Phys. Lett.}}, vol. 118, no.~22, p. 220501, May 2021. [Online]. Available:
  \url{https://aip.scitation.org/doi/10.1063/5.0050117}
\BIBentrySTDinterwordspacing

\bibitem{bogaerts_silicon_2012}
\BIBentryALTinterwordspacing
W.~Bogaerts \emph{et~al.}, ``\BIBforeignlanguage{en}{Silicon microring
  resonators},'' \emph{\BIBforeignlanguage{en}{Laser \& Photon. Rev.}}, vol.~6,
  no.~1, pp. 47--73, Jan. 2012. [Online]. Available:
  \url{https://onlinelibrary.wiley.com/doi/10.1002/lpor.201100017}
\BIBentrySTDinterwordspacing

\bibitem{rios_controlled_2018}
\BIBentryALTinterwordspacing
C.~Rios \emph{et~al.}, ``\BIBforeignlanguage{en}{Controlled switching of
  phase-change materials by evanescent-field coupling in integrated photonics
  [{Invited}]},'' \emph{\BIBforeignlanguage{en}{Opt. Mater. Express}}, vol.~8,
  no.~9, p. 2455, Sep. 2018. [Online]. Available:
  \url{https://opg.optica.org/abstract.cfm?URI=ome-8-9-2455}
\BIBentrySTDinterwordspacing

\bibitem{xiaoLargescaleEnergyefficientTensorized2021}
\BIBentryALTinterwordspacing
X.~Xiao \emph{et~al.}, ``Large-scale and energy-efficient tensorized optical
  neural networks on iii--v-on-silicon moscap platform,'' \emph{APL Photonics},
  vol.~6, no.~12, p. 126107, 2021. [Online]. Available:
  \url{https://doi.org/10.1063/5.0070913}
\BIBentrySTDinterwordspacing

\bibitem{sackesyn_experimental_2021}
\BIBentryALTinterwordspacing
S.~Sackesyn \emph{et~al.}, ``\BIBforeignlanguage{EN}{Experimental realization
  of integrated photonic reservoir computing for nonlinear fiber distortion
  compensation},'' \emph{\BIBforeignlanguage{EN}{Opt. Express, OE}}, vol.~29,
  no.~20, pp. 30\,991--30\,997, Sep. 2021, publisher: Optica Publishing Group.
  [Online]. Available:
  \url{https://opg.optica.org/oe/abstract.cfm?uri=oe-29-20-30991}
\BIBentrySTDinterwordspacing

\bibitem{Miscuglio:2020aa}
M.~Miscuglio \emph{et~al.}, ``Artificial synapse with mnemonic functionality
  using gsst-based photonic integrated memory,'' in \emph{2020 International
  Applied Computational Electromagnetics Society Symposium (ACES)}, 2020, pp.
  1--3.

\bibitem{10.3389/fphy.2022.1017714}
\BIBentryALTinterwordspacing
K.~Mekemeza-Ona \emph{et~al.}, ``All optical q-switched laser based spiking
  neuron,'' \emph{Frontiers in Physics}, vol.~10, 2022. [Online]. Available:
  \url{https://www.frontiersin.org/articles/10.3389/fphy.2022.1017714}
\BIBentrySTDinterwordspacing

\bibitem{8203770}
Y.~Liu \emph{et~al.}, ``Fault injection attack on deep neural network,'' in
  \emph{2017 IEEE/ACM International Conference on Computer-Aided Design
  (ICCAD)}, 2017, pp. 131--138.

\bibitem{breier2018deeplaser}
J.~Breier \emph{et~al.}, ``Deeplaser: Practical fault attack on deep neural
  networks,'' 2018.

\bibitem{8013784}
C.~Torres-Huitzil and B.~Girau, ``Fault and error tolerance in neural networks:
  A review,'' \emph{IEEE Access}, vol.~5, pp. 17\,322--17\,341, 2017.

\bibitem{8645906}
B.~Salami \emph{et~al.}, ``On the resilience of rtl nn accelerators: Fault
  characterization and mitigation,'' in \emph{2018 30th International Symposium
  on Computer Architecture and High Performance Computing (SBAC-PAD)}, 2018,
  pp. 322--329.

\bibitem{9926241}
G.~Li \emph{et~al.}, ``Understanding error propagation in deep learning neural
  network (dnn) accelerators and applications,'' in \emph{SC17: International
  Conference for High Performance Computing, Networking, Storage and Analysis},
  2017, pp. 1--12.

\bibitem{8465834}
B.~Reagen \emph{et~al.}, ``Ares: A framework for quantifying the resilience of
  deep neural networks,'' in \emph{2018 55th ACM/ESDA/IEEE Design Automation
  Conference (DAC)}, 2018, pp. 1--6.

\bibitem{8053727}
E.~I. Vatajelu and L.~Anghel, ``Fully-connected single-layer stt-mtj-based
  spiking neural network under process variability,'' in \emph{2017 IEEE/ACM
  International Symposium on Nanoscale Architectures (NANOARCH)}, 2017, pp.
  21--26.

\bibitem{8758653}
E.-I. Vatajelu \emph{et~al.}, ``Special session: Reliability of
  hardware-implemented spiking neural networks (snn),'' in \emph{2019 IEEE 37th
  VLSI Test Symposium (VTS)}, 2019, pp. 1--8.

\bibitem{8875270}
A.~Bosio \emph{et~al.}, ``Rebooting computing: The challenges for test and
  reliability,'' in \emph{2019 IEEE International Symposium on Defect and Fault
  Tolerance in VLSI and Nanotechnology Systems (DFT)}, 2019, pp. 8138--8143.

\bibitem{9774711}
T.~Spyrou \emph{et~al.}, ``Reliability analysis of a spiking neural network
  hardware accelerator,'' in \emph{2022 Design, Automation \& Test in Europe
  Conference \& Exhibition (DATE)}, 2022, pp. 370--375.

\end{thebibliography}
\end{document}